\documentclass[12pt]{article}

\usepackage{amsmath, euscript, amssymb, amsfonts, latexsym}

\usepackage{mathrsfs}

\newcommand{\mW}{{\mathscr{W}}}
\newcommand{\mF}{{\mathscr{F}}}

\newcommand{\oR}{{\mathbb R}}
\newcommand{\oC}{{\mathbb C}}

\newcommand{\twast}{\mathbin{\circledast}}

\def\dd{{\delta^{(2)}}}

\newcommand{\tr}{\mathop{\mathrm{tr}}\nolimits}

\renewcommand{\Re}{\mathop{\mathrm{Re}}\nolimits}
\renewcommand{\Im}{\mathop{\mathrm{Im}}\nolimits}

\begin{document}

\baselineskip=20pt

\vspace{1.5cm}

\begin{center}
{\large\bf Integral representations of the star product corresponding to the $s$-ordering of the creation and annihilation operators}
%\footnote{This is an author-created, un-copyedited version of an article accepted for publication in  Physica Scripta. The publisher is not responsible for any errors or omissions in this version of the manuscript. The Version of Record is available online at https://doi.org/10.1088/0031-8949/90/7/074008}

\baselineskip=15pt

\vspace{0.8cm}

{\bf M.~A.~Soloviev}\footnote{E-mail: soloviev@lpi.ru}
\end{center}

\centerline{\small\sl I.~E.~Tamm Department of Theoretical Physics, P.~N.~Lebedev Physical Institute,}

 %\vspace{0.05cm}
 \centerline{\small\sl Russian Academy of Sciences,  Leninsky Prospect~53, 119991 Moscow, Russia}

\vskip 2em

\vspace{0.2cm}

\begin{abstract}
A new integral representation is obtained for the star product corresponding to the  $s$-ordering of the creation and annihilation operators. This parametric ordering convention introduced by Cahill and Glauber enables one  to vary the type of ordering in a continuous way from normal order to antinormal order.  Our  derivation  of the corresponding integral representation  is based on using reproducing formulas for analytic and antianalytic functions. We also discuss a different representation whose kernel is a generalized function  and compare the properties of this kernel with those of the kernels of another family of star products which are intermediate  between the  $qp-$ and $pq$-quantization.

\medskip
\end{abstract}

\noindent Keywords: deformation quantization, star product algebras, Weyl-Wigner correspondence, Wick and anti-Wick symbols

\section{\large Introduction}

\baselineskip=20pt

The star product corresponding to the Cahill-Glauber $s$-ordering  rule~\cite{CG} was first considered by Man'ko {\it et al}~\cite{MMM}, along with the star product corresponding to the symplectic tomography map. An integral representation for this star product was also  discussed in article~\cite{MMV}  devoted to the duality symmetry of star products and in recent review~\cite{LV} of the matrix bases for a variety of noncommutative star products based on a generalized Weyl-Wigner transform. As the order parameter $s$ changes from 1 to  -1, the  $s$-ordering of the creation and annihilation operators changes continuously from normal order to  antinormal order and becomes symmetric at $s=0$. This  parametrization has been proposed in~\cite{CG} as a useful tool  for understanding and comparison of the convergence properties of the expansions of operator functions in ordered powers of  the annihilation and creation operators for  these three principal forms of ordering.  The  integral representation derived below for the  star product corresponding to the $s$-ordering generalizes the well-known representation~\cite{BS} for the Wick star product corresponding to normal ordering and turns continuously into the latter  as $s\to +1$. Our method of deriving this representation is based on using reproducing formulas for analytic and antianalytic functions. A simple proof of these formulas  is given in Sec.~3 for the reader's convenience. It should be emphasized that our representation of the star product $f\star_s g$, where $f$ and $g$ are symbols of $s$-ordered operators, as well as the representation~\cite{BS} of the Wick star product, is expressed not through the functions $f$ and $g$ on the real phase space, but  through their analytic continuations to the complex space. An important point is that the corresponding integral kernel is absolutely integrable. In Sec.~5, we show that a representation of $f\star_s g$ with integration over the real phase space and with a bounded integrable kernel is possible only for  $s=0$, i.e., only for the Weyl-Moyal star product. If $s=\pm 1$, then the integral kernel of such a representation cannot be an ordinary function and is not even a tempered distribution. However, for any $s$, it can be interpreted  as a singular generalized function defined on a suitable space of analytic test functions, which is characterized precisely in the same section. When written in this form, the integral kernel depends continuously on $s$. For a comparison, we derive in Sec.~6 an integral representation for another family of star products corresponding to  a parametrized $t$-ordering of the coordinate and momentum operators, which  is intermediate between standard and antistandard ordering or, in other words, between  $p\,q$ and  $qp$-quantization. This representation is not new, see, e.g.,~\cite{BD}, where it was derived in a different way. Its  kernel is bounded and locally integrable on the real phase space for all $-1<t<+1$ and, as we shall see, turns continuously into  a tempered distribution as $t \to\pm 1$.

A few words about notation may be useful. For simplicity, we  consider the case of two-dimensional phase space with canonically conjugate coordinates $q$ and $p$ whose Poisson bracket is equal to unity.
The complex coordinates $z$ and   $\bar z$ are given by
\begin{equation}
z=\frac{1}{\sqrt2}(\varsigma q+i\varsigma^{-1}p),\quad \bar
z=\frac{1}{\sqrt2}(\varsigma q-i\varsigma^{-1}p),
\notag
\end{equation}
where $\varsigma$ is a dimensional parameter such that  $\varsigma q$ and $\varsigma^{-1}p$ have the same dimension. We will use  two Fourier transformations  $\mathcal F\colon f\to\tilde f$ and $\breve{\mathcal F}\colon f\to\breve f$ defined by
\begin{align}
\tilde  f(w,\bar w)&=\frac{1}{\pi}\int e^{z\bar w-\bar z
w}f(z,\bar z) d^2z \label{1.1}\\\intertext{and} \breve  f(w,\bar w)&=\frac{1}{\pi}\int e^{\bar z w-z\bar w}f(z,\bar z) d^2z,
\label{1.2}
\end{align}
where $d^2z=d(\Re z) d(\Im z)$ is a real element of area in the complex plane. It is natural to call them symplectic Fourier transformations, because
\begin{equation}
i(\bar z w- z \bar w)=2\Im(z\bar w)
 \notag
\end{equation}
is  the canonical symplectic 2-form on the pase space.
In contrast to the ordinary Fourier transformation, both transformations $\mathcal F$ and $\breve{\mathcal F}$  are involutions,
\begin{equation}\mathcal F^2=\breve{\mathcal F^2}=\mathrm{Id}.
\notag
\end{equation}
We let $\dd(z)$ denote the two-dimensional delta-function $\delta(\Re z)\delta(\Im z)$. Clearly,
\begin{equation}
\dd(z)=\frac{1}{\pi^2}\int e^{z\bar w-\bar z w}d^2z.
 \notag
\end{equation}
As usual, $a$ and $a^\dagger$ are the annihilation and creation operators acting on a Hilbert space   and satisfying the commutation relation
 \begin{equation}
[a,a^\dagger]=\hbar.
 \label{1.3}
\end{equation}
These operators differ by a factor of  $\hbar^{1/2}$  from operators denoted by the same symbols in~\cite{CG}, whose commutator equals the unit operator. The operators used here  are preferable from the standpoint of  deformation quantization theory treating Planck's constant  $\hbar$ as a deformation parameter.  To each complex number  $w$  we assign the  operator
\begin{equation}
D(w)=e^{w a^\dagger-\bar w
a},
\label{1.4}
\end{equation}
called in~\cite{CG} displacement operator. It follows from~\eqref{1.3} and the Baker-Campbell-Hausdorff formula that the operators~\eqref{1.4} satisfy the relation
\begin{equation}
D(w)D(w')=e^{i\hbar\Im(w\bar w')}D(w+w').
\label{1.5}
\end{equation}
In other words, these operators realize a unitary projective representation of  the translation group of phase space  with the multiplier  $e^{i\hbar\Im(w\bar w')}$.

\section{\large The $s$-ordering and its corresponding star product}

The Weyl correspondence between phase space functions and  operators acting on a Hilbert space can be written as
 \begin{equation}
f \longmapsto A_f=\frac{1}{\pi}\int \tilde f(w,\bar w)D(w)d^2w.
 \label{2.1}
\end{equation}
 The monomial  $\bar z^\alpha z^\beta$ is transformed by the mapping~\eqref{2.1} into the symmetrically ordered product  $\{(a^\dagger)^\alpha a^\beta\}$. For instance, $\{a^\dagger a\}=\tfrac12(a^\dagger a+ aa^\dagger)$. The
$s$-ordered product of the operators $a$ and $a^\dagger$ emerges if~\eqref{2.1} is replaced by the correspondence
\begin{equation}
f \longmapsto A_{f,s}=\frac{1}{\pi}\int \tilde f(w,\bar w)D_s(w)d^2w,
 \label{2.2}
\end{equation}
where
\begin{equation}
D_s(w)= e^{\tfrac{\hbar s}{2} |w|^2} D(w).
\label{2.3}
\end{equation}
By using the formula
\begin{equation}
\widetilde{\bar z^\alpha
z^\beta}=\pi(-1)^\alpha\partial^\alpha_w\partial^\beta_{\bar
w} \delta^{(2)}(w),
\notag
\end{equation}
we find that the monomial $\bar z^\alpha z^\beta$ is transformed by~\eqref{2.2} into the operator
\begin{equation}
\{(a^\dagger)^\alpha a^\beta\}_s =\left.
\frac{\partial^{\alpha+\beta}D_s(w)}{{\partial w^\alpha \partial(-\bar w)^\beta}}\right|_{w=0}.
 \label{2.4}
\end{equation}
This is precisely the definition of  $s$-ordered product given in~\cite{CG}. From the equalities
\begin{equation}
D_1(w)= e^{w a^\dagger}e^{-\bar w a}\quad\text{and} \quad D_{-1}(w)=e^{-\bar w a} e^{w a^\dagger},
\notag
\end{equation}
it is clear that the orderings specified by $s=+1,-1$ are, respectively, normal and aninormal,
\begin{equation}
\{(a^\dagger)^\alpha a^\beta\}_1=(a^\dagger)^\alpha a^\beta,\quad
\{(a^\dagger)^\alpha a^\beta\}_{-1}= a^\beta (a^\dagger)^\alpha.
\notag
\end{equation}
The star product $f\star_s g$ corresponding to the  $s$-ordering or, in other words, the composition law  for phase space functions that is induced by the operator product through  the correspondence~\eqref{2.2} can be obtained by a direct generalization of the method used by von Neumann~\cite{N} for the Weyl correspondence. In~\cite{S13}, such a derivation has been carried out  for orderings of a more general form, and for the reader's convenience, we reproduce it in notation used here. First we show that in terms of the Fourier transforms the required composition law is given by
\begin{equation}
(\tilde f\twast_s\tilde g)(w,\bar w)=\frac{1}{\pi}\int \tilde f(w-w',\bar
w-\bar w')\tilde g(w',\bar w')
e^{\tfrac{\hbar}{2}\,\left[(1-s)(w- w') \bar w'-(1+s)(\bar w-\bar w')
w'\right]}d^2w'.
 \label{2.5}
\end{equation}
We note that the integral on the right-hand side of~\eqref{2.5}  is a noncommutative deformation of convolution and turns into the ordinary convolution as $\hbar\to0$. If  $s=0$, the exponential on the right-hand side of~\eqref{2.5} takes the form $e^{i\hbar\Im(w\bar w')}$ and, in this case, the deformed convolution is called  "twisted convolution".  Substituting $\tilde f\twast_s\tilde g$ for $\tilde f$ in~\eqref{2.2}, using the definition~\eqref{2.3} and changing integration variables, we find that
\begin{multline}
\int (\tilde f\twast_s\tilde g)(w,\bar w)D_s(w)d^2w=
\int (\tilde f\twast_s\tilde g)(w,\bar w)e^{\tfrac{\hbar s}{2}|w|^2}D(w)d^2w=\\=
\frac{1}{\pi}\iint \tilde f(w,\bar
w)\tilde g(w',\bar w')
e^{\tfrac{\hbar}{2}\,\left[\,(1-s)w \bar w'-(1+s)\bar w
w'+ s|w+w'|^2 \right]}D(w+w')d^2wd^2w'.
\notag
\end{multline}
The expression in square brackets can be rewritten as
\begin{equation}
s|w|^2+s|w'|^2+2i\Im(w\bar w').
 \notag
\end{equation}
By using~\eqref{1.5}, we obtain
\begin{multline}
\frac{1}{\pi}\int (\tilde f\twast_s\tilde g)(w,\bar w)D_s(w)d^2w=\\=
\frac{1}{\pi^2}\iint \tilde f(w,\bar w)\tilde g(w',\bar w')D_s(w)D_s(w')d^2wd^2w'=
A_{f,s}A_{g,s},
 \notag
\end{multline}
which is what we set out to prove. Now we define $f\star_s g$ by
\begin{equation}
f\star_s g=\mathcal F(\tilde f\twast_s\tilde g)
 \label{2.6}
\end{equation}
and conclude that the generalized Weyl transformation~\eqref{2.2} converts this star product into the operator product. An explicit expression for $f\star_s g$ follows directly from the definitions~\eqref{2.5}, \eqref{2.6}. Taking the Fourier transform of the deformed convolution~\eqref{2.5}, we find that
\begin{multline}
(f\star_ s g)(z,\bar z)= \frac{1}{\pi}\int (\tilde f\twast_s\tilde g)(w,\bar w)e^{w\bar z-\bar w, z}d^2w=\\= \frac{1}{\pi^2}\iint\tilde f(w,\bar w)\tilde g(w',\bar
w') e^{\tfrac{\hbar}{2}\,\left[(1-s)w
\bar w'-(1+s)\bar w\cdot w'\right]+(w+w') \bar z- (\bar w+\bar w')z}d^2w d^2w'=\\
 =\frac{1}{\pi}\int \tilde  f(w,\bar w) g\left(z-\tfrac{\hbar}{2}(1-s)w,\bar z-\tfrac{\hbar}{2}(1+s)\bar w\right)
e^{w\bar z- \bar w z}d^2w.
 \label{2.7}
\end{multline}
On substituting the explicit expression for $\tilde  f$, we obtain the representation
\begin{multline}
(f\star_ s g)(z,\bar z)=\\=\frac{1}{\pi^2}\iint
f(z',\bar z')g\left(z-\tfrac{\hbar}{2}(1-s)w,\bar z-\tfrac{\hbar}{2}(1+s)\bar w\right)
 e^{w(\bar z-\bar z')- \bar w(z-z')}d^2z'd^2w.
 \label{2.8}
\end{multline}
 By changing the integration variables from $w$ and $\bar w$  to $z''=z-\tfrac{\hbar}{2}(1-s)w$ and $\bar z''=\bar z-\tfrac{\hbar}{2}(1+s)\bar w$, the star product can be given the form
\begin{equation}
(f\star_s g)(z, \bar z)=\iint f(z',\bar z')g(z'',\bar z'') k_s(z'-z,z''-z)d^2z'd^2z'',
 \label{2.9}
\end{equation}
where
\begin{equation}
 k_s(z',z'')=\frac{4}{(\pi\hbar)^2(1-s^2)}
 \exp\left\{\frac{2}{\hbar}\left(\frac{\bar z'z''}{1-s}-\frac{z'\bar z''}{1+s}\right)\right\}.
 \label{2.10}
\end{equation}
The integral kernel~\eqref{2.10} was first obtained  by Man'ko {\it et al}~\cite{MMM} by a different calculation, see Remark in Sec.~5. A formal proof of this representation is also presented in~\cite{BD}, appendix A.5. If $s=0$, then the exponent in~\eqref{2.10} is pure imaginary and we have
\begin{equation}
 k_0(z',z'')=\frac{4}{(\pi\hbar)^2}
 \exp\left\{-\frac{4i}{\hbar}\Im(z'\bar z'')\right\}.
 \label{2.11}
\end{equation}
When expressed in terms of the real variables
$q$ and  $p$,  $k_0(z'-z,z''-z)$ coincides with the well-known expression for the integral kernel of the composition law for the Weyl symbols, see, e.g.,~\cite{BS}.

For $s\ne 0$, the real part of exponent in~\eqref{2.10}  is nonzero and is not negative definite. Therefore, the integral~\eqref{2.9} needs a careful interpretation. This question was analyzed in~\cite{LV}, where a prescription of how this integration should be performed has been given for $-1<s\le0$.  In particular, the integrals on the right-hand sides of~\eqref{2.8} and  \eqref{2.9} are not absolutely convergent for polynomial functions $f$ and $g$. In this case,  we can obtain the correct result by using instead~\eqref{2.7} and regarding $\tilde f$ as a tempered distribution. On the other hand, it is well known that the Wick star product, which corresponds to $s=1$, can be represented in the form of an integral over the complex plane, which is absolutely convergent not only for polynomial symbols  but also for all entire functions of order less than  2. Because the integral kernel~\eqref{2.10} is singular for $s=\pm1$, it is desirable to find  a  representation having the  absolute convergence property for an arbitrary $s$. To this end, we shall use reproducing  formulas for analytic and antianalytic functions, whose derivation is given in the next section.

\section{\large Reproducing formulas}

We start with the following easily verifiable identity
\begin{equation}
\frac{1}{\pi}\int
e^{z\bar v+\bar z w-z\bar z} d^2z=e^{\bar v w},\qquad v,w\in\oC.
 \label{3.1}
\end{equation}
Using~\eqref{3.1}, we obtain
\begin{equation}
\frac{1}{\pi}\int z^\alpha
e^{(w-z)\bar z} d^2z=\left.\frac{1}{\pi}\frac{\partial^\alpha}{\partial \bar v^\alpha}\int e^{z\bar v+\bar z w-z\bar z} d^2z\right|_{\bar v=0}= \left.\frac{\partial^\alpha e^{\bar v w}}{\partial \bar v^\alpha}\right|_{\bar v=0}= w^\alpha.
 \label{3.2}
\end{equation}
It follows that
\begin{equation}
\varphi(w)=\frac{1}{\pi}\int \varphi(z)
e^{(w-z)\bar z} d^2z
 \label{3.3}
\end{equation}
for any entire analytic function $\varphi(z)=\sum_\alpha a_\alpha z^\alpha$, whose Taylor coefficients satisfy the condition
\begin{equation}
\sum_\alpha \alpha!|a_\alpha|^2<+\infty.
 \label{3.4}
\end{equation}
Indeed, if~\eqref{3.4} is satisfied, then the Schwarz inequality yields
\begin{equation}
\left|\sum_\alpha a_\alpha z^\alpha\right|\le\left(\sum_\alpha \alpha!|a_\alpha|^2\right)^{1/2} \left(\sum_\alpha|z|^{2\alpha}/\alpha!\right)^{1/2}\le Ce^{|z|^2/2}.
 \notag
\end{equation}
This estimate enables us to interchange the order of summation and integration in the expression $\int\sum_\alpha a_\alpha z^\alpha e^{(w-z)\bar z} d^2z$ by the dominant convergence theorem, and then~\eqref{3.2} implies~\eqref{3.3}. Analogously, by differentiating~\eqref{3.1} with respect to $w$, we obtain the reproducing formula
\begin{equation}
\varphi(\bar v)=\frac{1}{\pi}\int \varphi(\bar z)
e^{(\bar v-\bar z) z} d^2z
 \label{3.5}
\end{equation}
for antianalytic functions, i.e., for entire functions  of the variable~$\bar z$. Furthermore, we have
\begin{equation}
\frac{1}{\pi}\int z^\alpha\bar z^\beta e^{-z\bar z}d^2z=
\left.\frac{\partial^{\alpha+\beta}}{\partial \bar v^\alpha\partial w^\beta}\int e^{z\bar v+\bar z w-z\bar z} d^2z\right|_{\substack{\bar v=0\\w=0}}=\left.\frac{\partial^{\alpha+\beta}e^{\bar v w}}{\partial \bar v^\alpha\partial w^\beta}\right|_{\substack{\bar v=0\\w=0}}=
\alpha!\delta_{\alpha\beta},
\label{3.6}
\end{equation}
where  $\delta_{\alpha\beta}$ is the Kronecker symbol.
The space of entire functions with the property~\eqref{3.4} can be made into a Hilbert space by giving it the scalar product $\langle\phi,\psi\rangle=\sum_\alpha \alpha!\,a_\alpha\bar b_\alpha$. The set of functions $z^\alpha/\sqrt{\alpha!}$, where $\alpha$ runs through all nonnegative integers, is clearly an orthonormal basis for this space, and~\eqref{3.6}  shows that its scalar product can be written as
\begin{equation}
\langle\varphi,\psi\rangle=\frac{1}{\pi}\int \varphi(z)\overline{\psi(z)}e^{-z\bar z}d^2z.
 \label{3.7}
\end{equation}
The functions $\bar z^\alpha/\sqrt{\alpha!}$ form  a basis for the analogous space of antianalytic functions.  Both these spaces are known as Fock-Bargmann spaces. In the context of deformation quantization, it is  useful to include $\hbar$  in the definitions and, in particular, to replace~\eqref{3.7} by the scalar product
\begin{equation}
\langle\varphi,\psi\rangle=\frac{1}{\pi\hbar}\int \varphi(z)\overline{\psi(z)}e^{-\tfrac{1}{\hbar}z\bar z}d^2z.
 \label{3.8}
\end{equation}
We let $\mF_\hbar$  denote the corresponding function space and, for its elements, we have the reproducing identity
\begin{equation}
\varphi(w)=\frac{1}{\pi\hbar}\int \varphi(z)
e^{\tfrac{1}{\hbar}(w-z)\bar z} d^2z.
\label{3.9}
\end{equation}
An analogous identity holds for the  space $\overline{\mF}_\hbar$ of antianalytic functions. It follows, from what has been said, that we also have the reproducing formula
\begin{equation}
\varphi(w,\bar w)=\frac{1}{(\pi\hbar)^2}\iint \varphi(z, \bar z')
e^{\tfrac{1}{\hbar}(w-z)\bar z+\tfrac{1}{\hbar}(\bar w-\bar z') z'} d^2zd^2z',
 \label{3.10}
\end{equation}
which holds for all functions belonging to the Hilbert tensor product $\mF_\hbar\otimes\overline{\mF}_\hbar$.

\section{\large An integral  representation of the star product $f\star_sg$}

We now return to  the last integral in~\eqref{2.7}. Considering the second function in the integrand  as an analytic function of  $w$ and $\bar w$ for fixed $z$ and applying the reproducing formula~\eqref{3.10}, we may write it in the form
\begin{multline}
g\left(z-\tfrac{\hbar}{2}(1-s)w,\bar z-\tfrac{\hbar}{2}(1+s)\bar w\right)=\\=
\frac{1}{(\pi\hbar)^2}\iint g(z+z',\bar z+\bar z'')
e^{\tfrac{1}{\hbar}\left(-\tfrac{\hbar}{2}(1-s)w-z'\right)\bar z'+\tfrac{1}{\hbar}\left(-\tfrac{\hbar}{2}(1+s)\bar w-\bar z''\right) z''} d^2z'd^2z''
 \label{4.1}
\end{multline}
Substituting this expression into~\eqref{2.7} and integrating over  $w$, we obtain the representation
\begin{multline}
(f\star_s g)(z,\bar z)=\frac{1}{(\pi\hbar)^2}\iint f\left(z+\tfrac12(1+s)z'',\bar
z-\tfrac12(1-s)\bar z'\right) g(z+z',\bar z
+\bar z'')\\
\times e^{-\tfrac{1}{\hbar}\left(z'\bar z'+z''\bar
z''\right)}d^2z'd^2z''.
 \label{4.2}
\end{multline}
For any $s$, the integral~\eqref{4.2} is absolutely convergent for polynomials. Moreover, it converges absolutely for all entire functions of order $<2$.

If $s=1$, we can integrate over $z'$ by  using~\eqref{3.9} and reduce the double integral to a single integral. It is clear, however, that in this  case, the same result is obtained at once  by  applying~\eqref{3.5} to the function $g(z,\bar z-\hbar\bar w)$. Then changing the integration variable, we reproduce the well-known integral representation~\cite{BS} for the Wick star product
\begin{equation}
(f\star_{W} g)(z,\bar z)=\frac{1}{\pi\hbar}\int f(z', \bar z)g(z,\bar z')e^{-\tfrac{1}{\hbar}(z'-z)(\bar z'-\bar z)}d^2z'.
\label{4.3}
\end{equation}
There is also no need to use~\eqref{3.10} in the case of the anti-Wick star product $(f\star_{\rm AW} g)$ specified by $s=-1$ and corresponding to the antinormal ordering. Applying~\eqref{3.9} to $g(z-\hbar w,\bar z)$, we obtain
\begin{equation}
(f\star_{AW} g)(z,\bar z)=\frac{1}{\pi\hbar}\int f(z, \bar z-\bar z')g(z+z',\bar z)e^{-\tfrac{1}{\hbar}z'\bar z'}d^2z'.
\label{4.4}
\end{equation}
Let us  show that this representation is in full agreement with the usual representation~~\cite{BS}  of the anti-Wick star product in the deferential form
\begin{equation}
(f\star_{AW} g)(z,\bar z)=f(z,\bar
z)\,e^{-\hbar\,
 \overleftarrow{\partial_{\bar z}}
 \,\overrightarrow{\partial_z}}g(z,\bar z).
 \label{4.5}
\end{equation}
It suffices to demonstrate this  for   monomials. Using~\eqref{4.5}, we find that
\begin{equation}
(z^\alpha\bar z^{\bar\alpha})\star_{AW}(z^\beta \bar z^{\bar\beta})
=z^\alpha \bar z^{\bar\beta}\sum_{\kappa=0}^{\min(\bar\alpha,\beta)}\frac{(-\hbar)^\kappa}{\kappa!}
\frac{\bar\alpha!}{(\bar\alpha-\kappa)!}\frac{\beta!}{(\beta-\kappa)!}\bar
z^{\bar\alpha-\kappa}z^{\beta-\kappa}. \notag
\end{equation}
On the other hand,  the representation~\eqref{4.4} yields
\begin{multline}
(z^\alpha\bar z^{\bar\alpha})\star_{AW}(z^\beta \bar z^{\bar\beta})=\frac{1}{\pi\hbar}\int z^\alpha(\bar z-\bar z')^{\bar\alpha}(z+z')^\beta  \bar z^{\bar\beta}
e^{-\frac{1}{\hbar}z'\bar z'}d^2z'=\\=
z^\alpha \bar z^{\bar\beta}\sum_{\kappa=0}^{\bar\alpha}\sum_{\lambda=0}^\beta
\frac{\bar\alpha!}{\kappa!(\bar\alpha-\kappa)!}\frac{\beta!}{\lambda!(\beta-\lambda)!}
\bar z^{\bar\alpha-\kappa}z^{\beta-\lambda}(-1)^\kappa \frac{1}{\pi\hbar}\int\bar
z^{\prime^\kappa} {z'}^\lambda e^{-\frac{1}{\hbar}z'\bar z'}d^2z'.
 \notag
\end{multline}
We therefore obtain the same result by virtue of the orthogonality relation~\eqref{3.6}.

To demonstrate the efficiency of  representation~\eqref{4.2}, we calculate the $s$-star product of two Gaussian functions
$g_a=e^{-az\bar z}$  and  $g_b=e^{-bz\bar z}$. Substituting them into~\eqref{4.2}, setting $s_-=\frac12(1-s)$, $s_+=\frac12(1+s)$ for notational convenience, and first performing the integration over $z'$ with the use of~\eqref{3.1}, we obtain
\begin{multline}
(g_a\star_s g_b)(z,\bar z)=\frac{1}{(\pi\hbar)^2}\iint e^{-a(z+s_+z'')(\bar z-s_-\bar z')-b (z+z')(\bar z+\bar z'')}
 e^{-\tfrac{1}{\hbar}\left(z'\bar z'+z''\bar
z''\right)}d^2z'd^2z''=\\=
\frac{1}{\pi\hbar}e^{-(a+b)z\bar z}\int e^{-as_+z''\bar z-b z\bar z''} e^{-\tfrac{1}{\hbar}z''\bar
z''} e^{-\hbar abs_-(z+s_+z'')(\bar z-\bar z'')}d^2z''.
 \notag
\end{multline}
Setting $c_s=1/(1+\hbar^2abs_-s_+)$ and  using~\eqref{3.1} again, but this time  with $w=-b(1+\hbar as_-)z$ and $\bar v=-as_+(1+\hbar bs_-)\bar z$, we find that
\begin{equation}
(g_a\star_s g_b)(z,\bar z)=c_s \exp\{-(a+b+\hbar abs_-)z\bar z+c_s\hbar abs_+(1+\hbar as_-)(1+\hbar bs_-)z\bar z\}.
 \notag
\end{equation}
After substituting the explicit form of $c_s$, the exponent becomes
\begin{equation}
-\frac{a+b+\hbar ab(s_--s_+)}{1+\hbar^2abs_-s_+}z\bar z,
  \notag
\end{equation}
and we conclude that
\begin{equation}
(g_a\star_s g_b)(z,\bar z)=\frac{4}{4+\hbar^2ab(1-s^2)} \exp\left\{-\frac{4(a+b-\hbar abs)}{4+\hbar^2ab(1-s^2)}z\bar z\right\}.
 \label{4.6}
\end{equation}
For $s=0$, \eqref{4.6} turns into the well-known formula for the Weyl-Moyal star product of two Gaussians, see, e.g, Eqs.~(48) and (49) in \cite{DN},  where $2/a^2$ and $2/b^2$ correspond to our $a$ and~$b$.

\section{\large Integral kernels as generalized functions}

In this section, we derive another representation of the star product  $f\star_sg$ by using the inverse of the transformation~\eqref{2.2}. This inverse mapping plays the same role as the Wigner mapping for the Weyl correspondence~\eqref{2.1}. We will also use the formula for the trace of the displacement operator
\begin{equation}
\tr D(w)=\pi\delta^2(w),
 \label{5.1}
\end{equation}
which is derived by employing the basic properties of  coherent states, see~\cite{CG,P}. The transformation~\eqref{2.2} can be rewritten as
\begin{equation}
f \longmapsto A_{f,s}=\frac{1}{\pi}\int  f(z,\bar z)\breve D_s(z))d^2z,
 \label{5.2}
\end{equation}
where the symplectic Fourier transform $\breve D_s$ is defined by~\eqref{1.2}.  It follows from~\eqref{1.5}, \eqref{2.3}, and~\eqref{5.1} that
\begin{multline}
\tr(\breve D_s(z)\breve D_{-s}(z'))=\frac{1}{\pi^2}\iint  e^{z\cdot\bar w-\bar z\cdot
w+z'\cdot\bar w'-\bar z'\cdot w'}e^{s\hbar(|w|^2-|w'|^2)/2} \tr(D(w)D(w'))d^2wd^2w'=\\=
\frac{1}{\pi}\int  e^{(z-z')\cdot\bar w-(\bar z-\bar z')\cdot
w}d^2w=\pi\delta^{(2)}(z-z').
\notag
\end{multline}
Therefore the inverse of the transformation~\eqref{5.2} is
\begin{equation}
A\longmapsto f_{A,s}(z)=\tr(A \breve D_{-s}(z))
 \label{5.3}
\end{equation}
 The star product $f\star_sg$ is obtained by applying the mapping~\eqref{5.3} to the operator $A_{f,s}A_{g,s}$, and hence  we have
 \begin{equation}
(f\star_s g)(z,\bar z)=\tr(A_{f,s}A_{g,s}\breve D_{-s}(z)).
 \label{5.4}
\end{equation}
Writing it formally as an integral over the phase space,
\begin{equation}
(f\star_sg)(z)=\iint f(z',\bar z')g(z'',\bar z'')K_s(z', z'', z)d^2z'd^2z'',
 \label{5.5}
\end{equation}
we see that the integral kernel   $K_s$ is given by
\begin{gather}
K_s(z',z'',z)=\frac{1}{\pi^2}\tr(\breve D_s(z')\breve D_s(z'')\breve D_{-s}(z))\label{*}\\=
\frac{1}{\pi^5}\iiint e^{2i\Im(z'\bar w'+z''\bar w''+z\bar w)}\tr(D_s(w')D_s(w'')D_{-s}(w))d^2w'd^2w''d^2w.
\label{5.6}
\end{gather}
The relation~\eqref{1.5} implies that
\begin{equation}
D(w)D(w')D(w'')=e^{i\hbar\Im(w\bar w'+w\bar w''+w'\bar w'')}D(w+w'+w'').
\notag
\end{equation}
Hence
\begin{multline}
\tr(D_s(w')D_s(w'')D_{-s}(w))=\\=e^{\hbar s(|w'|^2+|w''|^2-|w|^2)/2}
e^{i\hbar\Im(w'\bar w''+w'\bar w+w''\bar w)}
\tr D(w'+w''+w)=\\=
\pi e^{-\hbar s\Re(w'\bar w'')+i\hbar\Im(w'\bar w'')}\delta^{(2)}(w'+w''+w).
\label{5.7}
\end{multline}
Substituting the last expression into~\eqref{5.6}, we find that  $K(z',z'',z)$ can be written as
\begin{equation}
K_s(z',z'',z)=k_s(z'-z,z''-z),
\label{5.8}
\end{equation}
where
\begin{equation}
 k_s(z',z'')=\frac{1}{\pi^2}\,\mathcal F_{\prime}\mathcal  F_{\prime\prime}(e^{-\hbar s\Re(w'\bar w'')+i\hbar\Im(w'\bar w'')}).
\label{5.9}
\end{equation}
For $s=0$, we have
\begin{equation}
\mathcal F_{\prime}\mathcal  F_{\prime\prime}(e^{i\hbar\Im(w'\bar w'')})=4\hbar^{-2}e^{-4i\hbar^{-1}\Im (z'\bar z'')}
\notag
\end{equation}
and arrive again at the formula~\eqref{2.11}.

If $s\ne 0$, then the real part of the exponent in~\eqref{5.9}, expressed in  terms of the real variables, is an indefinite quadratic form with  signature $(+,+,-,-)$. Therefore the double Fourier transform of this exponential function is in general a singular generalized function. The corresponding test function space, on which it is certainly well defined, can  be described by using the elementary inequality  $|\Re(w'\bar w'')|\le (|w'|^2+|w''|^2)/2$. Let $w'=(u'+iv')/\sqrt2$ and  $w''=(u''+iv'')/\sqrt2$. Clearly, the function  $e^{-\hbar s\Re(w'\bar w'')}$ becomes integrable after multiplication by any function of the form
\begin{equation}
\varphi(u',v',u'',v'') e^{-\hbar |s|(u'^2+v'^2+u''^2+v''^2)/4},
  \label{5.10}
\end{equation}
where $\varphi$ belongs to the Schwartz space $S(\oR^4)$  of smooth rapidly decreasing functions. Let $\mathcal W_{2,\hbar |s|/4}(\oR^4)$ be the space of all function of the form~\eqref{5.10}. In the one-variable case, the elements of $\mathcal W_{2,a}(\oR)$ are characterized by the inequalities
\begin{equation}
|\partial^\alpha g(u)|\le
C_{\alpha,N}(1+|u|)^{-N}e^{-a u^2},\qquad \alpha,
N=0,1,2,\dots.
  \notag
\end{equation}
In the case of several variables, the space $\mathcal W_{2,a}$ is defined by the same formula but with $\alpha$ considered as a multi-index.  The Fourier transformation maps this space onto the space  $\mathcal W^{\,2,b}$, where $b=(4a)^{-1}$, whose elements admit  analytic continuation to  entire functions satisfying the estimate
\begin{equation}
|(x+iy)^\alpha f(x+iy)|\le
C_\alpha e^{b\, y^2}.
  \notag
\end{equation}
It is  easy to see that the generalized function~\eqref{5.9} can be written as
\begin{equation}
k_s(z',z'')=e^{\tfrac{\hbar}{2}\left[(s+1)\partial_{z'}
 \partial_{\bar z''}+ (s-1)\partial_{\bar z'}
 \partial_{z''}\right]}\delta^{(2)}(z')\delta^{(2)}(z'').
 \label{5.11}
 \end{equation}
Indeed, taking the double Fourier transform of the right-hand side of~\eqref{5.11}, we obtain
\begin{multline}
\frac{1}{\pi^2}\iint e^{z'\bar w'-\bar z' w'+z''\bar w''-\bar z'' w''}e^{\tfrac{\hbar}{2}\left[(s+1)\partial_{z'}
 \partial_{\bar z''}+ (s-1)\partial_{\bar z'}
 \partial_{z''}\right]}\delta^{(2)}(z')\delta^{(2)}(z'')d^2z'd^2z''=\\=
 \frac{1}{\pi^2}e^{-\tfrac{\hbar}{2}\left[(s+1)\bar w'w''+(s-1)w'\bar w''\right]}=\frac{1}{\pi^2}e^{-\hbar s\Re(w'\bar w'')+i\hbar\Im(w'\bar w'')}.
 \notag
 \end{multline}
The generalized function~\eqref{5.11} is well defined on  $\mathcal W^{\,2,b}$ with $b=(\hbar |s|)^{-1}$, because the infinite order differential operator $e^{\tfrac{\hbar}{2}\left[(s+1)\partial_{z'}
 \partial_{\bar z''}+ (s-1)\partial_{\bar z'}
 \partial_{z''}\right]}$ is well defined on this space and maps it continuously into $S$. More particularly, applying this operator to a function in  $\mathcal W^{\,2,1/\hbar s}(\oR^4)$ yields a series which converges absolutely  in $S(\oR^4)$. Substituting~\eqref{5.8} and \eqref{5.11} into~\eqref{5.5}, we obtain the representation
\begin{equation}
(f\star_s g)(z,\bar z)=f(z,\bar
z)\,e^{\tfrac{\hbar}{2}\,\left[(s+1)\overleftarrow{\partial_z}
 \,\overrightarrow{\partial_{\bar z}}+
 (s-1)\overleftarrow{\partial_{\bar z}}
 \,\overrightarrow{\partial_z}\right]}g(z,\bar z),
 \label{5.12}
\end{equation}
which holds for all $f,g\in \mathcal W^{\,2,1/\hbar s}$. In a manner similar to that used in Sec.~4 for the anti-Wick product, it is easy to verify that~\eqref{4.2} and~\eqref{5.12} give the same expression for $(z^\alpha\bar z^{\bar\alpha})\star_s(z^\beta \bar z^{\bar\beta})$. It should be pointed out that the kernel~\eqref{5.11} is continuous in $s$ under the topology of the dual of $\mathcal W^{\,2,1/\hbar s}$. We also note that  multiplication by the function $e^{\hbar s|w|^2/2}$, where $s>0$, is converted by the Fourier transformation into the differential operator  $e^{-(\hbar s/2)\partial_z\partial_{\bar z}}$, which maps $\mathcal W^{\,2,1/\hbar s}(\oR^2)$ isomorphically onto  $S(\oR^2)$, and it follows from~\eqref{2.2} and \eqref{2.3}  that
\begin{equation}
e^{-\tfrac{\hbar s}{2}
\partial_z\partial_{\bar z}}(f\star_s g)=
(e^{-\tfrac{\hbar s}{2}
\partial_z\partial_{\bar z}}f)\star
(e^{-\tfrac{\hbar s}{2}
\partial_z\partial_{\bar z}}g),
 \label{5.13}
\end{equation}
where $\star$ is the Weyl-Moyal star product, and  $f$ and $g$  are any two functions in $\mathcal W^{\,2,1/\hbar s}$.

{\large Remark.} As shown in~\cite{CG}, the symplectic Fourier transform of the $s$-ordered displacement operator~\eqref{2.3} can be expressed in the form
\begin{equation}
\breve D_s(z)=\frac{2}{1-s}D(z)\left(\frac{s+1}{s-1}\right)^{a^\dagger a}D(-z).
 \label{5.14}
\end{equation}
The integral kernel~\eqref{2.10} was found in~\cite{MMM,LV} starting from~\eqref{*} and using the expression~\eqref{5.14} for $\breve D_s$. This implies that the representation~\eqref{2.9} with $k_s$ given by~\eqref{2.10} and properly defined integration must be equivalent to~\eqref{5.12} for functions in $\mathcal W^{\,2,1/\hbar s}$ or for functions in a dense subspace of $\mathcal W^{\,2,1/\hbar s}$.

\section{\large Another one-parameter family of orderings}

We now consider another parametric ordering, indexed by  a real parameter $t$, which plays for the $p\,q-$ and  $qp-$quantization the same role as the Cahill-Glauber ordering for the Wick and anti-Wick quantization. Let    $Q$ and  $P$ be the position and momentum operators satisfying the Heisenberg commutation relation
\begin{equation}
[Q,P]=i\hbar.
\notag
\end{equation}
The displacement operator is now given by
\begin{equation}
D(u,v)=e^{i(vQ-uP)},
\notag
\end{equation}
where $u$ and $v$ are real variables\footnote{These variables differ by a factor of $\sqrt{2}$ from those in Sec.~5. }. The relation~\eqref{1.5} takes the form
\begin{equation}
D(u,v)D(u',v')=e^{i\hbar(vu'-uv')/2}D(u+u',v+v'),
\label{6.1}
\end{equation}
Let us define the operator $\mathsf D_t(u,v)$ by
\begin{equation}
\mathsf D_t(u,v)=e^{i\hbar t uv/2}e^{i(vQ-uP)}
\notag
\end{equation}
and consider, instead of~\eqref{2.2}, the correspondence
\begin{equation}
f \longmapsto A_{f,t}=\frac{1}{2\pi}\int \tilde f(u,v)
\mathsf D_t(u,v)\,du\,dv,
 \label{6.2}
\end{equation}
where
\begin{equation}
\tilde f(u,v)=(\mathcal Ff)(u,v)=\frac{1}{2\pi}\int e^{i(pu-qv)}f(q,p)\, dq\,dp.
\notag
\end{equation}
The monomial $q^\alpha p^\beta$ is transformed by the mapping~\eqref{6.2} into the product
\begin{equation}
\left\lgroup Q^\alpha P^\beta\right\rgroup_{\!t} =\left.
\frac{\partial^{\alpha+\beta}\mathsf D_t(u,v)}{{\partial v^\alpha \partial(-u)^\beta}}\right|_{\substack{u=0\\v=0}}.
 \label{6.3}
\end{equation}
Since $\mathsf D_1= e^{v Q}e^{-uP}$ and $\mathsf D_{-1}=e^{-u P} e^{vQ}$, we see that the orderings specified by $t=+1,-1$ are, respectively, standard and antistandard,
\begin{equation}
\left\lgroup Q^\alpha P^\beta\right\rgroup_{\!\!1}=Q^\alpha P^\beta,\qquad\left\lgroup Q^\alpha P^\beta\right\rgroup_{\!\!-1}= P^\beta Q^\alpha.
\notag
\end{equation}
Clearly, the Weyl symmetric ordering is specified by $t=0$.

To find the star product that corresponds to the ordering~\eqref{6.3}, we proceed along the same lines as in Sec.~2 and first define a deformed convolution  $\tilde f\twast_t\tilde g$ by
\begin{equation}
(\tilde f\twast_t\tilde g)(u,v)=\frac{1}{2\pi}\int \tilde f(u-u',
v-\bar v')\tilde g(u',v')
e^{\tfrac{i\hbar}{2}\,\left[(1-t)(v-v')
u'-(1+t)(u- u')v'\right]}du'\,dv'.
 \label{6.4}
\end{equation}
By using~\eqref{6.1}, it is easy to see that
\begin{multline}
\frac{1}{2\pi}\int_{\oR^2} (\tilde f\twast_t\tilde g)(u,v)\mathsf D_t(u,v)du\,dv=\\=
\frac{1}{(2\pi)^2}\int_{\oR^4} \tilde f(u,v)\tilde g(u',v')\mathsf D_t(u,v)\mathsf D_t(u',v')\,du\,dv\,du'dv'=
A_{f,t}A_{g,t},
 \notag
\end{multline}
Now we define  $f\star_tg$ as the symplectic Fourier transform $\tilde f\twast_t\tilde g$ of and find that
\begin{multline}
(f\star_t g)(q,p)= \frac{1}{2\pi}\int_{\oR^2} (\tilde f\twast_t\tilde g)(u,v)e^{i(vq-up)}\,du\,dv=\\= \frac{1}{(2\pi)^2}\int_{\oR^4}\tilde f(u,v)\tilde g(u',v') e^{\tfrac{i\hbar}{2}\,\left[(1-t)v
u'-(1+t)u v'\right]+i\left[(v+v') q- (u+u')p\right]}du\,dv\,du'dv'=\\
 =\frac{1}{2\pi}\int_{\oR^2} \tilde  f(u,v) g\left(q-\tfrac{\hbar}{2}(1+t)u,p-\tfrac{\hbar}{2}(1-t)v\right)
e^{i(vq-up)}du\,dv.
 \notag
\end{multline}
After substituting  the explicit form of $\tilde  f(u,v)$, the last integral becomes
\begin{equation}
\frac{1}{2\pi}\int_{\oR^4}
f(q',p')g\left(q-\tfrac{\hbar}{2}(1+t)u,p-\tfrac{\hbar}{2}(1-t) v\right)
 e^{i\left[v(q-q')- u(p-p')\right]}dq'dp'du\,dv.
 \notag
\end{equation}
Changing the integration variables from $u$ and $v$ to $q''=q-\frac{\hbar}{2}(1+t)u$ and $p''=p-\frac{\hbar}{2}(1-t)v$, we obtain the  representation
\begin{equation}
(f\star_t g)(q,p)= \int_{\oR^4}
f(q',p')g(q'',p'')\,\mathsf K_t(q'-q,p'-p,q''-q,p''-p)\,
 dq'dp'dq''dp'',
 \label{6.5}
\end{equation}
where
\begin{equation}
\mathsf K_t(q',p',q'',p'')= \frac{1}{(\pi\hbar)^2|1-t^2|}\exp\left\{-\frac{2i}{\hbar}\left(
\frac{p'q''}{1+t}-\frac{q'p''}{1-t}\right)\right\}.
  \label{6.6}
\end{equation}
It can be verified that the same result is obtained by a computation analogous to that carried out in Sec.~5 or, to be more specific, by using the mapping
\begin{equation}
A\longmapsto f_{A,t}(q,p)=\tr(A \breve{\mathsf D}_{-t}(q,p)),
 \notag
\end{equation}
which is   inverse to the mapping~\eqref{6.2}. For $t=0$, the formula~\eqref{6.6}  is clearly equivalent to~\eqref{2.11}. As $t\to\pm1$, the expression on the right-hand side of~\eqref{6.6} becomes singular.
Changing to the variables $(q\pm p)/\sqrt2$ and using the method of stationary phase, it can be shown that
 \begin{equation}
\lim_{\tau\to+\infty}\tau e^{\pm 2i\tau\, qp}=\pi \delta(q)\delta(p).
 \notag
\end{equation}
Therefore,
\begin{align*}
\mathsf K_1(q',p',q'',p'')&= \frac{1}{2\pi\hbar} \, e^{-\tfrac{i}{\hbar}p'q''}\delta(q')\delta(p''),\\\intertext{and}
\mathsf K_{-1}(q',p',q'',p'')&= \frac{1}{2\pi\hbar} \, e^{\tfrac{i}{\hbar}q'p''}\delta(p')\delta(q'').
\end{align*}
Thus, the kernel of the integral representation~\eqref{6.5} for the star product corresponding to the ordering~\eqref{6.3} is an ordinary locally integrable  function for all $t\ne\pm1$ and contains delta-functions for $t=\pm1$, whereas in the case of the Cahill-Glauber $s$-ordering, the kernel of the analogous representation lies outside the space of tempered distributions for $s=\pm1$.

\section{\large Concluding remarks}
The main result of this paper is the integral representation~\eqref{4.2} with absolutely integrable kernel for the $s$-star product,  which connects continuously the Wick and anti-Wick star products in accordance with the initial Cahill and Glauber's idea of varying the type of ordering in a continuous way from normal order to antinormal order.
It is well known that if an operator  $A$ is bounded, then its Wick symbol is the restriction of an entire function $f_A(z,z')$ to the set  $z'=\bar z$. An analogous statement holds for the $s$-symbols
$f_{A,s}$ with any $s>0$. Because of this, the representation~\eqref{4.2} is most useful in the case of positive  $s$. The anti-Wick symbols of bounded operators are generally not well-behaved functions and not even tempered distributions. The same can be said with respect to the  $s$-symbols for   $s<0$, and in this case, the representation~\eqref{2.7} appears to be more relevant. The Weyl-Moyal star product holds a central position in deformation quantization because it is completely determined by the symplectic structure of the phase space. The Schwartz space $S$ is an algebra under this star product and Schwartz's theory of  distributions provides a convenient framework for the Weyl symbol calculus, see, e.g.,~\cite{Fol}.  However the space $S$ is not an algebra with respect to the star product~\eqref{5.12} for any $s\ne0$ and, in particular it is not an algebra under the Wick and ant-Wick products. In this situation, it is helpful to use, along with Schwartz's test functions or instead of them, the function space    $\mW^2=\bigcap\limits_{b>0}\mathcal W^{2,b}$. As shown in \cite{S13},  $\mW^2$ is an algebra with respect to any translation invariant star product, including those defined by~\eqref{5.12}. By using the spaces $\mW^2$ and $\mathcal W^{2,b}$, we obtain an exact characterization~\cite{S14} of the extensions of the Wick and anti-Wick correspondences that are in line with the extension of the Weyl correspondence to   distributions. A space analogous to $\mathcal W^{2,b}$ was considered in~\cite{AMP}, where it was denoted by $G^{\frac12,\sigma^2}$ and  used in the context of the coarse scale description of  dynamics in terms of a smoothed Wigner transform. The function space $\mW^2$ also arises  naturally in quantum field theory on noncommutative spaces and proved useful for analyzing the nonlocal effects and the causality properties of noncommutative models~\cite{S07}.

\section*{\small ACKNOWLEDGMENTS}  This paper was supported in part by the the Russian Foundation for Basic Research (Grant No.~12-01-00865)

\end{document}